# Generation of second-mode internal solitary waves during the winter by a shoaling internal tide in the northern South China Sea

Jian Jun Liang[1], Xiao-Ming Li [1,2,3]

[1]Key Laboratory of Digital Earth Science, Institute of Remote Sensing and Digital Earth, Chinese Academy of Sciences, Beijing, China

[2] Laboratory for Regional Oceanography and Numerical Modeling, Qingdao National Laboratory for Marine Science and Technology, Qingdao, 266235, China

[2]Hainan Key Laboratory of Earth Observation, Sanya, China

Corresponding author: Xiao-Ming Li  (lixm@radi.ac.cn)

**Key Points:**

- The evolution of an internal tide mode is examined with intermodal coupling.
- A generation mechanism of observed mode-2 internal solitary waves during the winter is presented.
- A mode-1 internal tide can efficiently scatter into mode-2 waves during the winter.

**Abstract**

Field measurements of second-mode internal solitary waves (mode-2 ISWs) during the winter on the upper continental slope of the northern South China Sea were reported in Yang et al. (2009), but their generation mechanism remains elusive. We investigated this issue with a multi-modal evolution model and theoretical analysis, which suggest that the observed mode-2 ISWs were generated by a shoaling mode-2 semidiurnal internal tide (IT) that emanated from the Luzon Strait. The results show that two groups of mode-2 ISWs usually appear within one semidiurnal tidal period, successively riding on expanded and subsequently compressed pycnoclines. The number of wave groups largely depends on the amplitudes of the ITs; that is, a larger IT produces larger mode-2 ISWs. Furthermore, intermodal coupling dominates the evolution of a mode-1 IT, highlighting the importance of considering mode scattering in the propagation of low-mode ITs.

**1 Introduction**

Internal solitary waves (ISWs) are nonlinear density perturbations propagating horizontally within a pycnocline in the ocean (Jackson et al., 2012). The characteristics of vertical structures are used to classify them into either first-mode (mode-1) ISWs or second-mode (mode-2) ISWs. Mode-1 ISWs that displace isopycnals downward through the water column are known as ISWs of depression, whereas those that displace isopycnals upward through the water column are known as ISWs of elevation (Liu et al., 1998). Similar to mode-1 ISWs, mode-2 ISWs are also categorized into two types: convex waves and concave waves (Yang et al., 2010). Convex ISWs feature a double-humped structure of isopycnals throughout the water column by displacing upper isopycnals upward and lower isopycnals downward. In contrast, an hourglass-shaped structure characterizes concave ISWs.

Several generation mechanisms have been identified for convex ISWs. These include: (a) an ISW of depression shoaling over the slope-shelf topography and entering the breaking instability stage (Helfrich & Melville, 1986), (b) an ISW of depression shoaling over the slope-shelf topography and experiencing disintegration before breaking (Lamb & Warn-Varns, 2015; Liu et al., 2013), (c) a concave ISW shoaling over the slope-shelf topography and undergoing polarity conversion (Guo & Chen., 2012), (d) an ISW of elevation interacting with a steep sill (Vlasenko & Hutter, 2001), (e) the nonlinear disintegration of a mode-2 internal tide (IT) along a flat bottom (Chen et al. 2014), (f) the impingement of an internal tidal beam on a pycnocline (which is emanating from critical bathymetry) from below, (Grisouard et al., 2011), (g) the resonant interaction of a steady barotropic flow with an isolated topography (Stastna & Peltier, 2005), (h) the intrusion of the whole head of a gravity current into a three-layer fluid (Mehat et al., 2011), (i) the lee-wave mechanism(Ramp et al., 2012) and (j) a mesoscale-catelyzed mechanism (Dong et al., 2016).

In the past decades, most studies have focused on mode-1 ISWs in the northern South China Sea (Guo and Chen, 2014; Alford et al., 2015). Mode-2 ISWs and particularly their dynamics during the winter are yet poorly understood (Jackson et al., 2012). To date, the most comprehensive observation of convex ISWs on the upper continental slope in the northern South China Sea has been reported by Yang et al. (2009). The mooring position was located at $21.6145°N$ and $17.283°E$ at a depth of 350 m. Although convex ISWs occur both during the summer and during the winter, such ISWs were obviously more frequent in occurrence and demonstrated larger wave amplitudes during the winter. Furthermore, Furthermore, field measurements showed that ISWs of depression dominate the field records and most convex ISWs appear following the appearance of ISWs of depression during the summer, and thus, an intuitive hypothesis for the generation of convex ISWs is that they are generated by shoaling ISWs of depression. The hypothesis has been recently simulated by Lamb & Warn-Varns (2015). However, ISWs of depression were rarely observed and most convex ISWs appeared without the appearance of ISWs of depression during the winter, indicating a different generation mechanism from the summer case. Although convex ISWs are dominant and prevalent in the field records, their generation mechanism is still not fully understood, and none of the reported generation mechanisms can effectively explain the observed facts reported in Yang et al. (2009). Therefore, we hypothesize that the winter convex ISWs observed by Yang et al. (2009) in the northern SCS were generated by a shoaling IT that powerfully originated from the Luzon Strait.

The remainder of this paper is organized as follows. The numerical model, which incorporates intermodal coupling in the case of strong coupling between the first and second mode, used to simulate the generation of convex ISWs is presented in Section 2. A theoretical prediction based on the weakly nonlinear Korteweg-de Vries (KdV) theory is demonstrated in Section 3. The results are provided in Section 4, after which a discussion is presented in Section 5, followed by the conclusions in Section 6.

**2 Model Description**

Intermodal coupling can play an important role in the evolution of weakly nonlinear internal waves, particularly second-mode nonlinear internal waves (Shroyer et al., 2010). Therefore, the multi-modal evolution equation set found in Sakai and Redekopp (2009), which contains the derivation of the set from the Euler equations to the final evolution equation set comprising the first and second modes, is used to simulate the evolution of a shoaling IT. Here, we present only the final results of the derivation of this equation set.

The evolution equation for the first mode is

$$U_{1t} + \left(c_1^2 Z_1\right)_x = -\sum_{i=1}^{2}\sum_{j=1}^{2}\left(a_{1ij}^u U_i U_j + b_{1ij}^u U_{ix} U_j\right)$$
$$+ \sum_{i=1}^{2} d_{1i} U_{ixxt} - \sum_{i=1}^{2} r_{1i} c_i^2 Z_i,$$
$$Z_{1i} + U_{1x} = -\sum_{i=1}^{2}\sum_{j=1}^{2}\left(a_{1ij}^\sigma \left(U_i Z_j\right)_x + b_{1ij}^\sigma U_{1x} Z_j\right)$$
$$- \sum_{i=1}^{2} s_{1i} U_i, \quad (1)$$

and the evolution equation for the second mode is

$$U_{2t} + \left(c_2^2 Z_2\right)_x = -\sum_{i=1}^{2}\sum_{j=1}^{2}\left(a_{2ij}^u U_i U_j + b_{2ij}^u U_{ix} U_j\right)$$
$$+ \sum_{i=1}^{2} d_{2i} U_{ixxt} - \sum_{i=1}^{2} r_{2i} c_i^2 Z_i,$$
$$Z_{2i} + U_{2x} = -\sum_{i=1}^{2}\sum_{j=1}^{2}\left(a_{2ij}^\sigma \left(U_i Z_j\right)_x + b_{2ij}^\sigma U_{2x} Z_j\right)$$
$$- \sum_{i=1}^{2} s_{2i} U_i, \quad (2)$$

where $x$ and $t$ are space and time variables, respectively, $U_n (n=1,2)$ is the modal amplitude of the horizontal velocity, $Z_n (n=1,2)$ is the modal amplitude of the isopycnal displacement, $c_n (n=1,2)$ is the linear phase speed for long waves, and the coefficients are defined by the following set of relations:

$$a_{nij}^{(u)} = \langle \phi_n' \phi_i' \phi_j' \rangle, \quad b_{nij}^{(u)} = \left\langle \frac{N^2}{c_j^2} \phi_n' \phi_i \phi_j \right\rangle,$$

$$a_{nij}^{(\sigma)} = \left\langle \frac{N^2}{c_n^2} \phi_n \phi_i' \phi_j' \right\rangle, \quad b_{nij}^{(\sigma)} = \left\langle \frac{N^2}{c_n^2} \phi_n' \phi_i \phi_j \right\rangle,$$

$$d_{ni} = \langle \phi_n \phi_i \rangle, \quad r_{ni} = \left\langle \phi_n' \frac{\partial \phi_i'}{\partial x} \right\rangle,$$

$$s_{ni} = \frac{c_n^2}{I_n} \left[ \phi_n' \phi_i' \right]_{z=-h_b} \frac{dh_b}{dx} + r_{ni}, \qquad (3)$$

where the superscript denotes the partial derivative with respect to $z$, $\langle \cdots \rangle = \frac{c_n^2}{I_n} \int_{-h_b}^{0} (\cdots) dz$, $I_n = \int_{-h_b}^{0} N^2 \phi_n^2 dz$, $N$ is the buoyancy frequency, $h_b$ is the bottom depth and $\phi_n (n=1,2)$ is the eigenfunction determined by the solution of the following eigenvalue problem:

$$\left( \phi_n' \right)^2 + \left( N^2 / c_n^2 \right) \phi_n = 0; \quad \phi_n(z=0) = \phi_n(z=-h_b) = 0, n=1,2. \qquad (4)$$

Equations (1) and (2) are transformed into dimensionless form and then solved numerically using a finite-difference method, namely, a 4th-order compact scheme in space and a 3rd-order Adams-Bashforth scheme in time. A 4th-order compact filter is applied to every 20-unit time step for de-aliasing and stabilization. The grid resolution is $dx = 2.6 \times 10^{-4}$, and the time step is $dt = 1.8 \times 10^{-5}$. The selection of a high-resolution configuration ensures the presence of sufficiently resolving steep, nonlinear fronts and oscillatory waves.

The experimental set up in this study is constructed to facilitate a comparison with the observations of convex ISWs in December from Yang et al. (2009). Bathymetry data from 117 °E to 117.85 °E along a channel at 21.625 °N were derived from the ETOPO1 Global Relief Model (2009) over a distance of approximately 90 km. Background temperature and salinity profiles for December were selected from the World Ocean Atlas (2013). The buoyancy frequency was calculated through the Gibbs equation of state (i.e., the Thermodynamic Equation of Seawater,

TEOS-10). The western boundary is located at 117.85 °E, where an IT is assumed to enter the channel in the following form:

$$\eta_n(x=0,z,t) = -A_n \phi_n(z) \sin(\omega t). \tag{5}$$

In this study, six representative types of ITs are examined (Table 1).

Table 1 List of experiments in this study

| Case | Amplitude $A_n(m)$ | Frequency $\omega(10^{-4}s^{-1})$ | Mode $(n)$ | Note |
|---|---|---|---|---|
| 1 | 30 | 1.454 | 2 | Semidiurnal mode-2 IT |
| 2 | 30 | 0.727 | 2 | Diurnal mode-2 IT |
| 3 | 30 | 1.454 | 1 | Diurnal mode-1 IT |
| 4 | 30 | 0.727 | 1 | Semidiurnal mode-1 IT |
| 5 | 10 | 1.454 | 2 | Small semidiurnal mode-2 IT |
| 6 | 50 | 1.454 | 2 | Large semidiurnal mode-2 IT |

## 3 Theoretical Prediction

When the intermodal coupling effect is weak, the KdV equation is well known as an appropriate physical model for predicting the behaviors of weakly nonlinear internal waves. Considering an inviscid and incompressible fluid while assuming the flow is two-dimensional and neglecting the Earth's rotation, Grimshaw (2002) and Grimshaw et al. (2007) showed that a variable-coefficient KdV equation for the mode $n$ can be written as

$$\frac{\partial Z_n}{\partial t} + c_n \frac{\partial Z_n}{\partial x} + \frac{c_n Q_{nx}}{2Q_n} + \alpha_n Z_n \frac{\partial Z_n}{\partial x} + \beta_n \frac{\partial^3 Z_n}{\partial x^3} = 0. \tag{6}$$

Here, the coefficient $Q_n(x)$ represents the linear magnification factor, which is defined so that $Q_n Z_n^2$ is the wave action flux density for linear long waves. The coefficients $\alpha_n(x)$ and $\beta_n(x)$ of the nonlinear and dispersive terms, respectively, are determined by the waveguide properties of a basic state. More specifically, in the Boussinesq approximation, $\alpha_n(x)$ and $\beta_n(x)$ are expressed as

$$\alpha_n = \frac{3c_n}{2} \frac{\int_{-h_b}^{0} (\phi_n')^3 dz}{\int_{-h_b}^{0} (\phi_n')^2 dz}, \beta_n = \frac{c_n}{2} \frac{\int_{-h_b}^{0} (\phi_n)^2 dz}{\int_{-h_b}^{0} (\phi_n')^2 dz}.$$

(7)

For a mode-1 ISW, the KdV equation predicts a depression-type ISW when $\alpha_1 < 0$ and an elevation-type ISW when $\alpha_1 > 0$. For a mode-2 ISW, the KdV equation predicts a convex ISW when $\alpha_2 < 0$ and a concave ISW when $\alpha_2 > 0$. Fig. 1 shows that both $\alpha_1$ and $\alpha_2$ keep negative in the whole channel, implying that only ISWs of depression and convex ISWs can be generated by a shoaling IT.

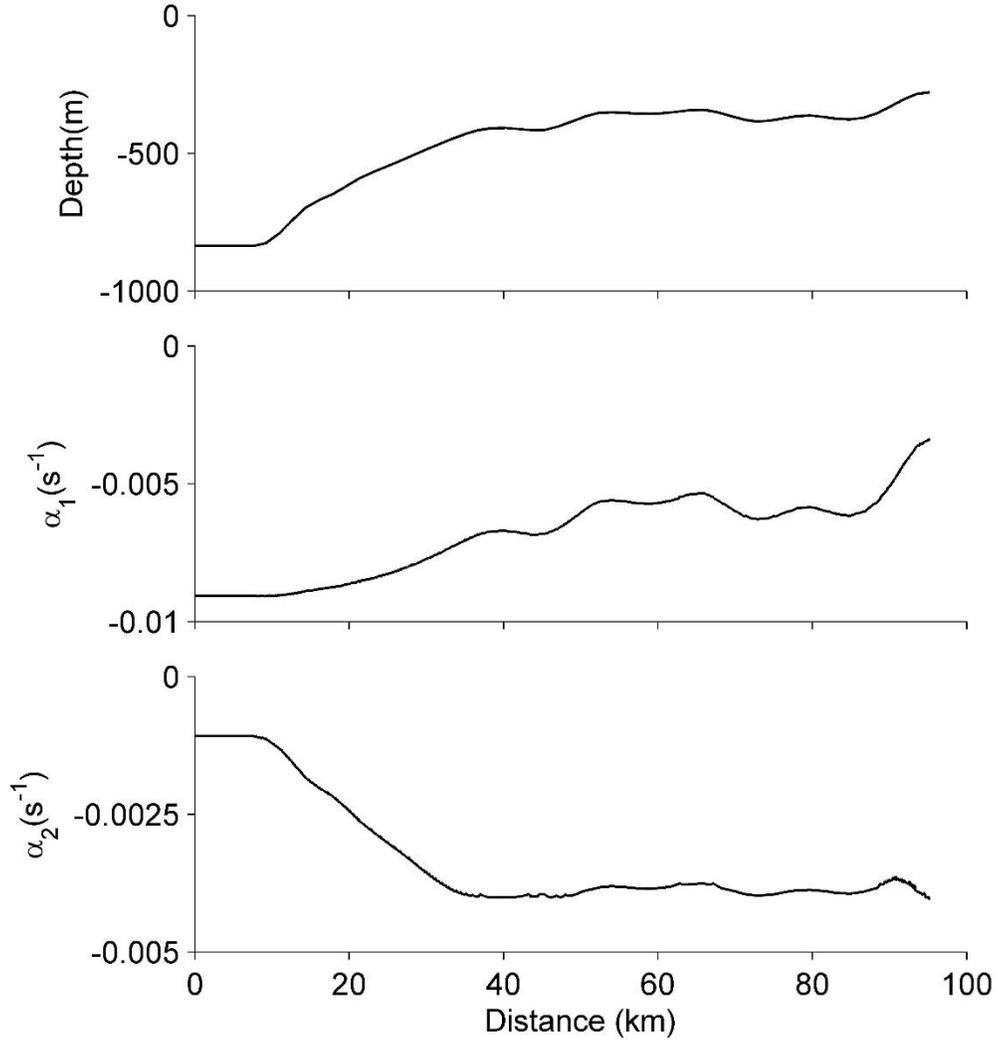

**Fig. 1 Nonlinear coefficients for the SCS during the winter plotted with the distance across a slope-shelf topography from an initial point at a depth of 837 m (the observations of Yang et al. (2009) were acquired at 350 m, and thus, the assumed IT originates from a greater depth)**

An ISW forms when the nonlinear wave steepening effect balances with linear wave dispersion. This balance occurs when a non-dimensional number, i.e., the Ursell number, which defines the ratio between nonlinearity and dispersion, satisfies $O(U_r) = O(\alpha A_{iw} L_{iw}^2 / \beta) = 1$, where $A_{iw}$ and $L_{iw}$ denote the wave amplitude and characteristic length, respectively (Li and Farmer, 2011). Therefore, a wave becomes steep when $U_r \gg 1$ or dispersive when $U_r \ll 1$. We calculated the values of $\log_{10} U_{r2}$ for a mode-2 internal wave with an amplitude of 30m and three typical wavelengths of 0.1 km, 1 km and 10 km, as depicted in Fig.2. From the figure, a shoaling mode-2 IT with an initial length of 10 km is predicted to undergo a strong nonlinear transformation and excite nonlinear waves with a length of 0.1km after a distance of 40 km, which provides theoretical support of the proposed hypothesis in this study.

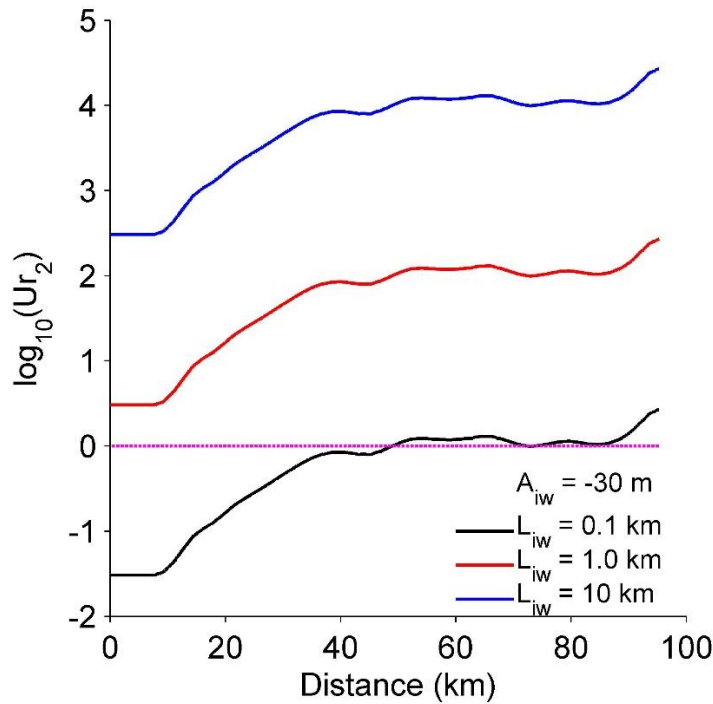

**Fig. 2 Ursell numbers for typical mode-2 internal waves during the winter as a function of distance. The critical Ursell number is indicated by the magenta dashed line**

## 4 Model Results

### 4.1 Generation process

The scenario of wave generation can be depicted by plotting the time evolution of the waveform. We use Case 1 to illustrate this issue. The maximum vertical displacements of the isopycnals in the first and second modes are shown in Fig. 3, from which it is evident that the form of an IT will be skewed under the effect of nonlinearity as the mode-2 IT shoals. The front face with positive slope becomes steep while the back face with a negative slope becomes gentle. A steep-faced shock (labelled A) is formed on the front face at a distance of approximately 39 km (a depth of 410 m). After further propagation, undulations develop and the shock is followed by a group of solitary depressions. Accompanying event A, a second shock (labelled B) is formed on the succeeding front face with a positive slope, bringing about a second group of solitary depressions. Thus, a shoaling semidiurnal mode-2 IT leads to the generation of two groups of solitary depressions.

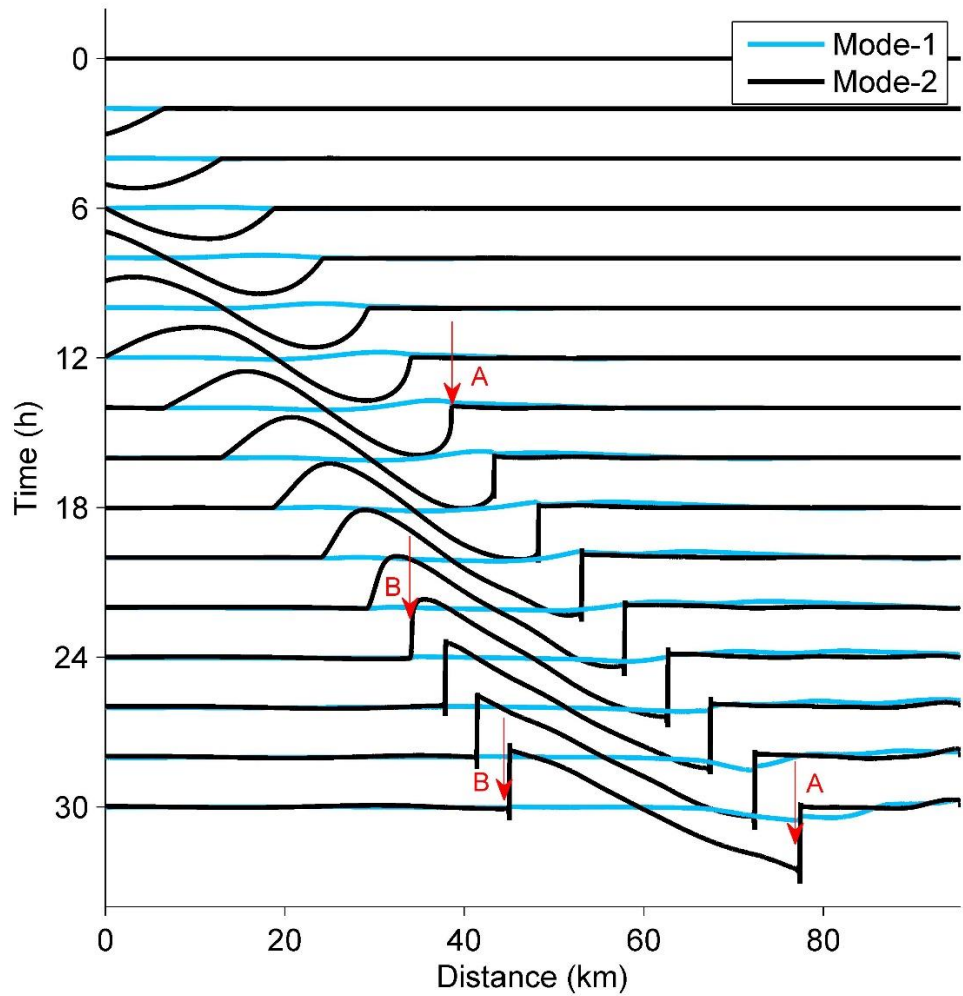

Fig. 3 The evolution of the maximum isopycnal displacements in the first and second modes for Case 1.

The vertical structures of the two groups of solitary depressions are shown in Fig. 4. Accordingly, the first group (group A) rides on an expanded pycnocline, whereas the second group (group B) rides on a compressed pycnocline. Moreover, magnified plots showing the waves in both groups (Fig. 5) depict a clear convex mode-2 wave structure with upward and anti-phase isopycnal displacements in the upper and lower layers, respectively (Fig. 5). The occurrence of convex type is consistent with the theoretical prediction in Section 3. Now, we make further considerations to confirm that the waves in the groups are convex ISWs. On the one hand, extracting the leading wave position of packet A from t=18h to t=30h yields that a leading wave propagating at an average speed of 0.67 m/s. Meanwhile, the average linear mode-2 phase speed for the spatial interval traversed by the leading wave is 0.56 m/s; after adding the nonlinear correction to this value, we obtain the nonlinear phase speed of 0.65m/s, which is in agreement with the phase speed obtained from the numerical results. On the other hand, the short waves in both groups have wavelengths of approximately 75 m, which is consistent with the order of $O(100\text{m})$ in predicted ISWs (shown in Fig. 2). Consequently, the agreement between the simulated results and the theory confirms that the waves in both groups are convex ISWs.

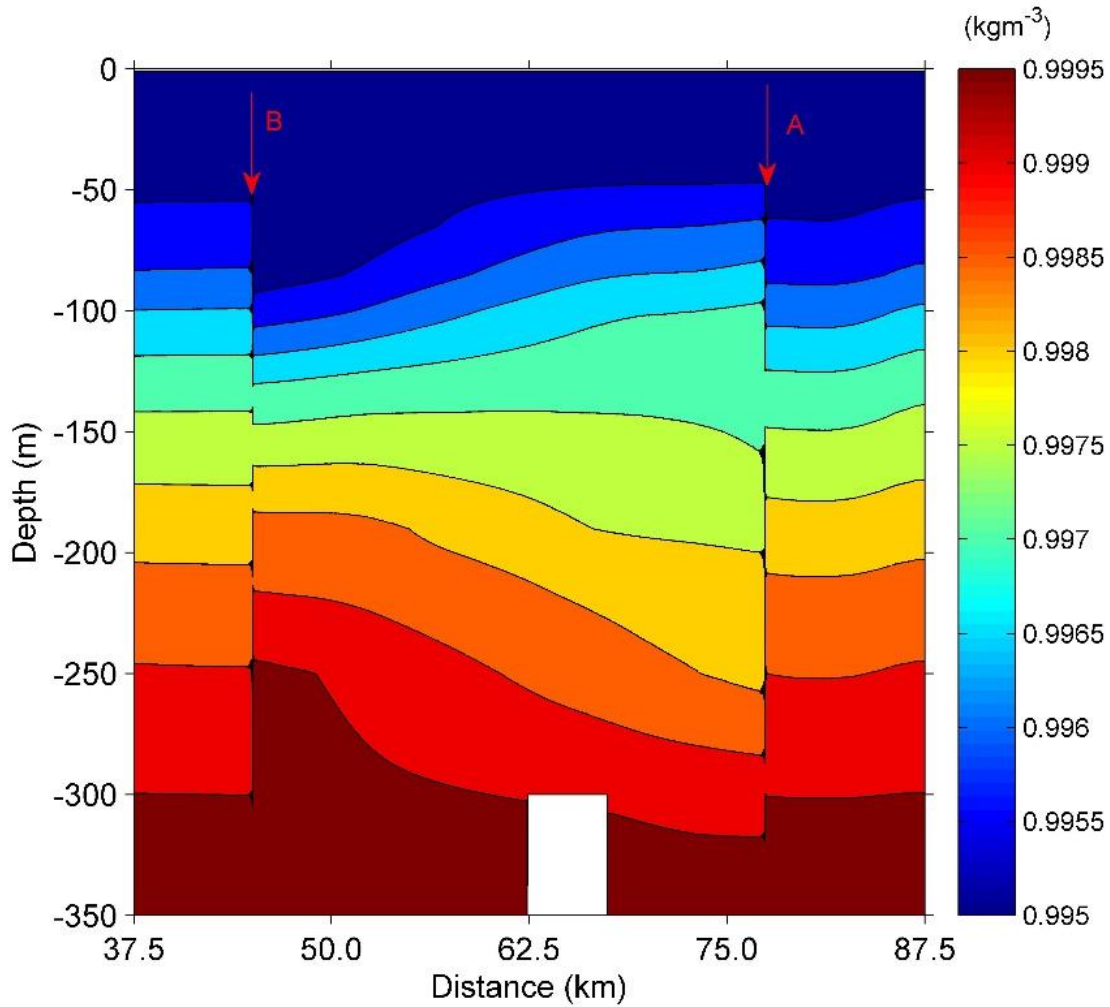

Fig. 4 Contour plots of the perturbation density divided by 1028 at t=30 h for Case 1

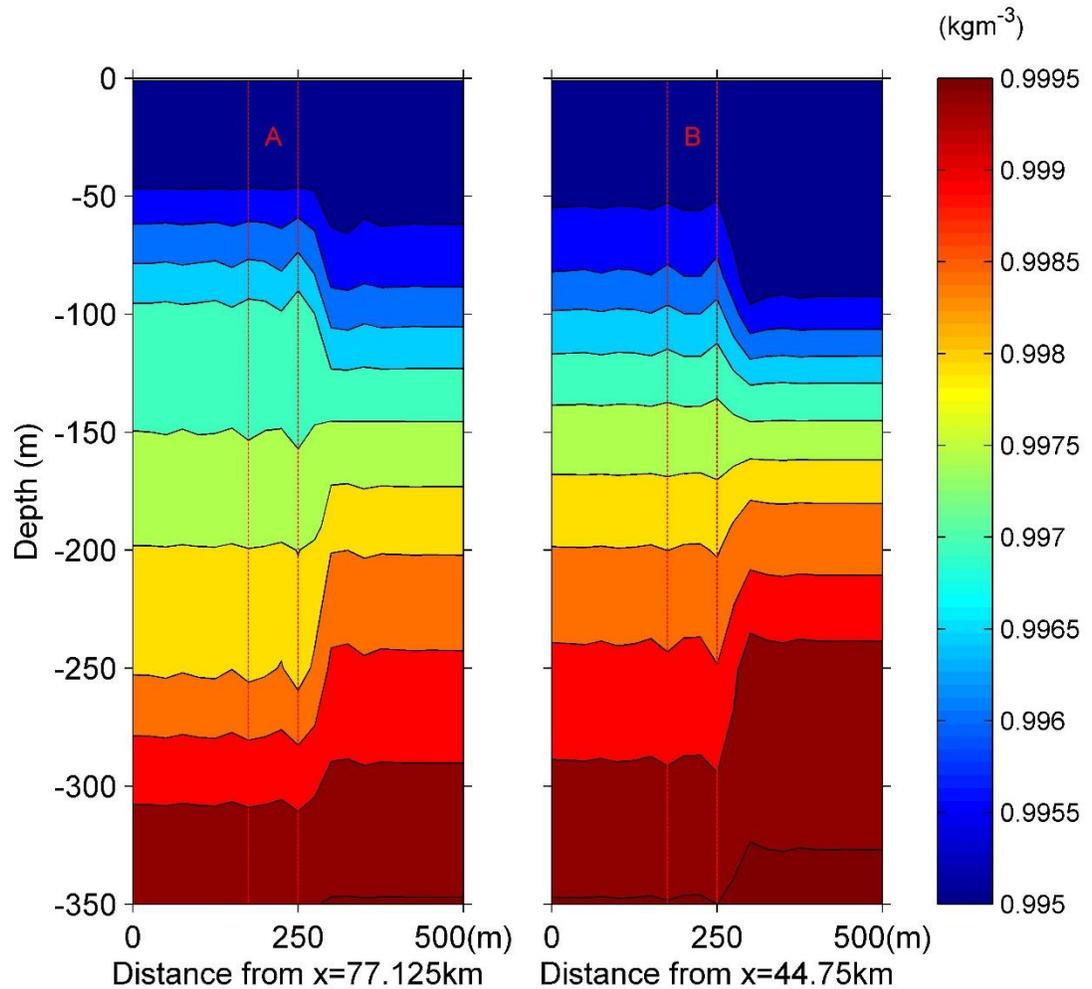

Fig. 5 Magnified view of groups A and B in Fig. 4

### 4.2 Effect of the waveform

To ensure that the convex ISWs observed during the winter by Yang et al. (2009) were generated by a shoaling mode-2 semidiurnal IT, we examined three other cases. Fig. 6 shows the wave fields at t=30 h for a shoaling mode-2 diurnal IT (upper panel), t=15 h for a shoaling mode-1 diurnal IT (middle panel), and t=15 h for a shoaling mode-1 semidiurnal IT (lower panel).

Remarkably distinct phenomena emerge in Fig.6. First, the shoaling mode-2 diurnal IT and the shoaling mode-1 semidiurnal IT each produce only one group of mode-2 and mode-1 ISWs, respectively, at approximately x=75 km. However, the shoaling mode-1 diurnal IT does not undergo nonlinear disintegration. Second, in contrast to mode-2 ITs, mode-1 ITs cannot propagate in a single mode; rather, they greatly experience strong intermodal coupling.

Moreover, efficient scattering into a mode-2 structure gives rise to a low-mode IT. Such a multi-modal structure, which is characteristic of a slanting wave front, is particularly evident at $x = 25\text{-}50$ km in the middle and lower panels. Third, although the shoaling mode-1 IT does not excite convex ISWs, it induces a nonlinear mode-2 wave at $x=50$ km, which represents a reflection of strong scattering from a mode-1 space to a mode-2 space.

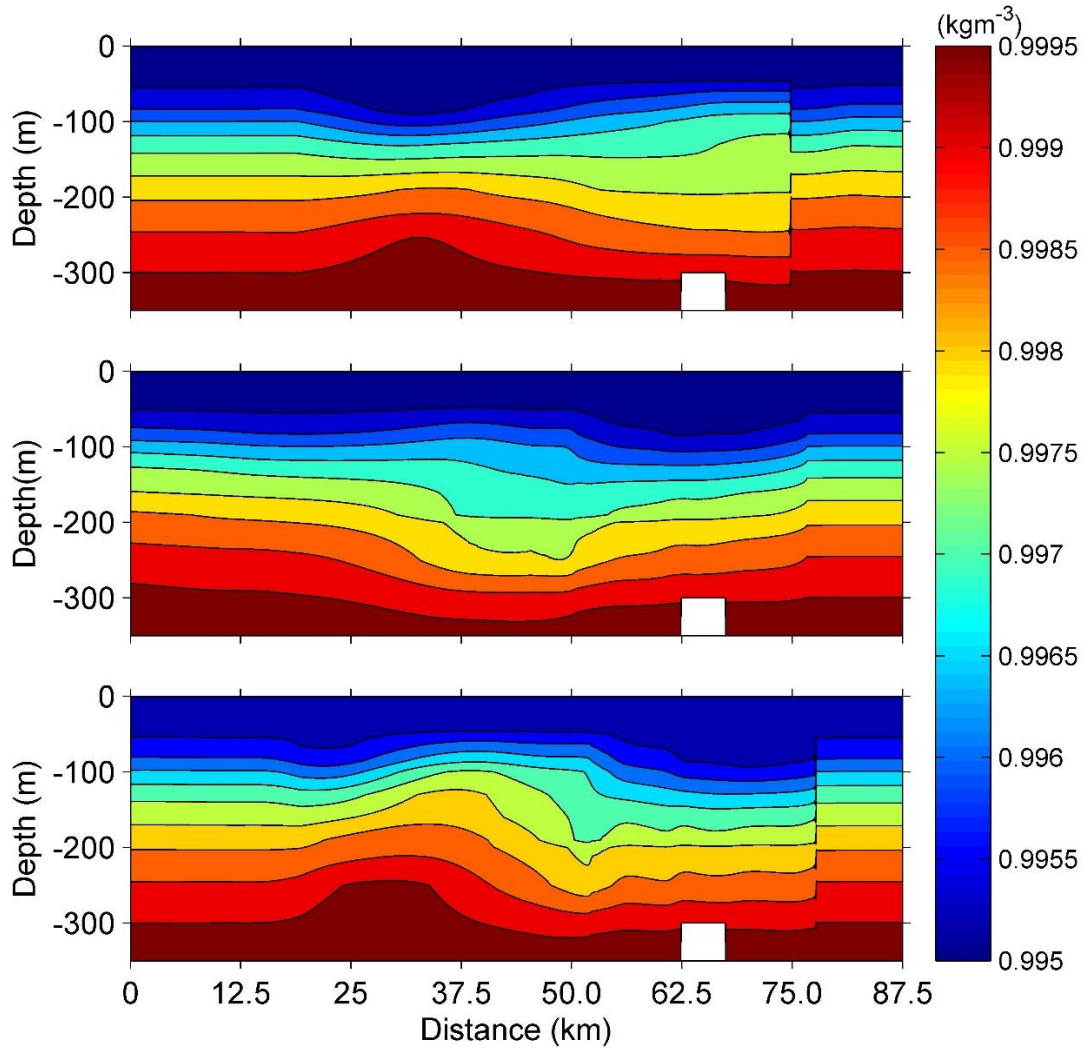

Fig. 6 Contour plots of the density divided by 1028 at t=30 h for Case 2 (upper panel) and t=15 h for Case 3 (middle panel) and Case 4 (lower panel)

### 4.3 Effect of the wave amplitude

The in-situ observations of Yang et al. (2009) indicate that a convex ISW may require a comparatively large IT. Therefore, this subsection is devoted to examining the influence of the IT amplitude. Fig. 7 contains the final wave fields for a shoaling mode-2 IT with amplitude of 50 m (upper panel) and 10 m (lower panel). As expected, a larger IT produces convex ISWs, whereas a smaller IT only propagates with a slightly nonlinear transformation. Moreover, an increase in the amplitude of a shoaling IT leads to additional groups of convex ISWs.

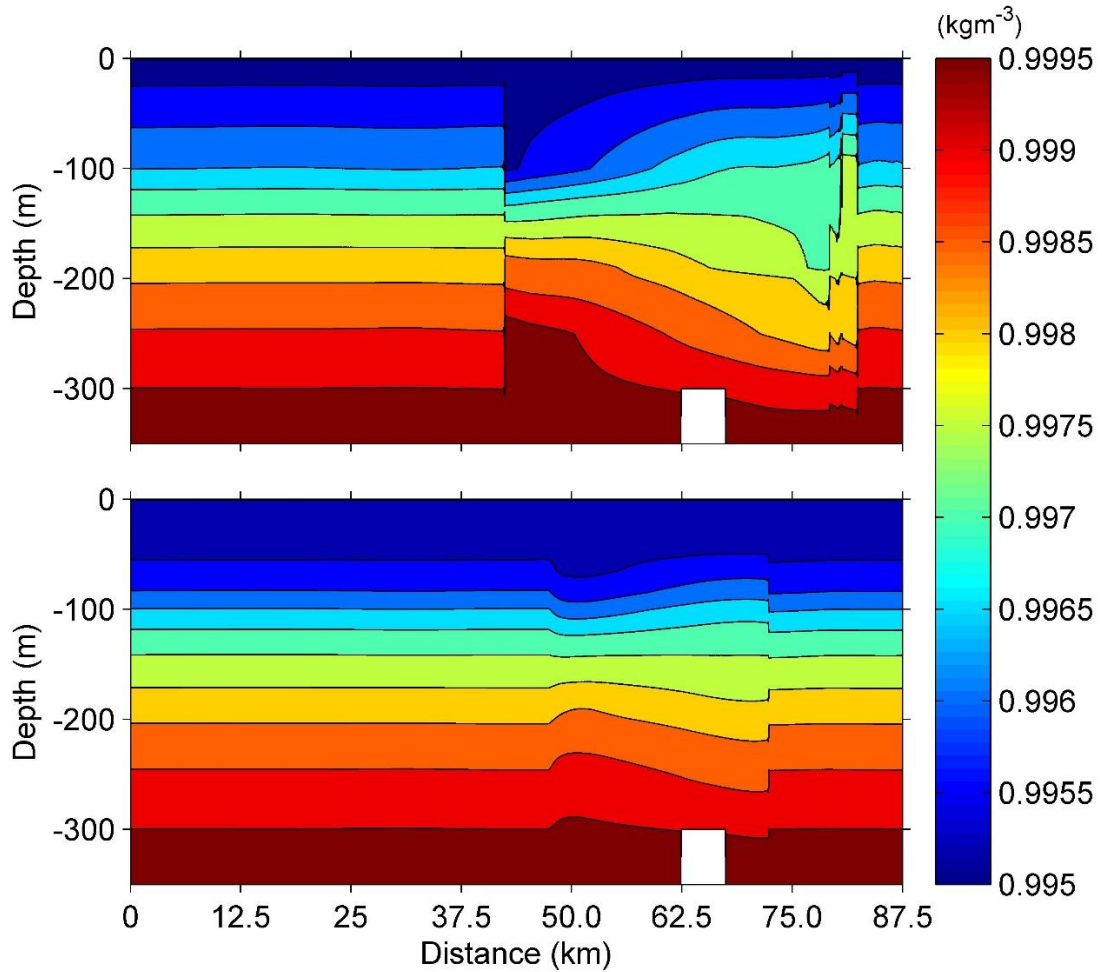

Fig. 7 Contour plots of the density divided by 1028 at t=30 h for Case 5 (upper panel) and Case 6 (lower panel)

## 5  Comparison between the Numerical Results and the Field Measurement

Two idealistic set-ups were adopted in the numerical simulation. One set-up uses the monthly averaged climatological stratification in December, which suffices to represent the mean stratification of the realistic ocean in December but cannot capture the variations on a smaller temporal scale such as caused by an eddy (Zhang et al, accepted for publication). The other set-up uses a sinusoidal tide to represent the incident internal tide mode at a depth of 830 m, an approach proposed and validated by Holloway et al. (1997; 1999). These two idealistic set-ups lead to some discrepancy between the simulated and observed wave profiles. However, scrutinizing the field measurements of convex ISWs by Yang et al. (2009) depicted in Fig. A1 in the Appendix A reveals the following four features that can coincide well with the numerical results presented above.

First, the consecutive arrival of waves on each day from 24 to 27 December 2005 and an indicative semidiurnal period of their arrival on each day from 24 to 26 December 2005 suggest an apparent tidal origin, supporting the proposed IT mechanism. The fact that they do not arrive at exactly the same time is probably due to the influence of mesoscale eddies and barotropic tidal currents on the propagation of ITs.

Second, Yang et al. (2009)'s measurements show that two wave groups usually arrived within a semidiurnal period, which are consistent with the waves A and B in Fig 3. In particular, the former wave group rode on an expanded pycnocline while the other wave group rode on a compressed pycnocline, in comparison to the background pycnocline. The unique wave location in the pycnocline was consistent with the numerical result depicted in Fig 4. Moreover, the second feature indicates that there was a second-mode semidiurnal IT in the field records, supporting the proposed second mode semidiurnal IT mechanism.

Third, an exceptional phenomenon occurred on 26 December 2005 in which three wave groups emerged within a semidiurnal period with the former two wave groups joined together. The phenomenon was reproduced in the upper panel in Fig. 7, corresponding to the modeling results of a 50-m second-mode semidiurnal IT.

Fourth, the isotherm displacement by the observed convex ISWs was $26 \pm 16$ m at a depth of 240 m (Yang et al., 2009), and the value was 13 m and 25 m at t=30 h in the 30-m and 50-m second-mode semidiurnal IT experiments, respectively. The simulated displacements were in the observed range.

By comparing the four shared features in the field measurements and our numerical simulations, we suggest that the observed convex ISWs are generated by a shoaling second mode semidiurnal IT.

## 6 Discussion

### 6.1 Origin of ITs

Numerical simulations confirm that convex ISWs during the winter can be produced by a strong shoaling semidiurnal IT. Accordingly, a straightforward question arises regarding where the required large amplitude IT originates.

We examined the origins of ITs by analyzing the barotropic forcing term following Baines (1982) that can be found as follows (da Silva et al. 2009):

$$F = zN^2 \int Q dt \cdot \nabla \left( -\frac{1}{h_b} \right) \quad (8)$$

where $Q$ is the barotropic mass flux vector $Q = (Q_x, Q_y) = (-u_b h_b, -v_b h_b)$, $u_b$ and $v_b$ denote the zonal and meridional components of the barotropic velocity, and the other symbols have the same meanings as in previous sections. We calculated the barotropic tidal forcing due to the $M_2$ constituent in the northern SCS because the convex ISWs originate from a semidiurnal IT. The barotropic tidal transport data were obtained from the Oregon State Tidal Inversion System (OTIS) Regional Tidal Solutions for the China Seas developed by Egbert and Erofeeva (2002) at a horizontal resolution of 1/30°. The temperature and salinity profiling data used to calculate the buoyancy frequency (N) were obtained from the World Ocean Atlas (2013) at a resolution of 1/4°. We acquired the bathymetry data from the 1/60° ETOPO1 Global Relief Model (2009).

A map of the maximum depth-integrated barotropic forcing term over a complete tidal cycle on 23 December 2005 is shown in Fig. 8, which reveals that the Luzon Strait is the most active area of barotropic tidal forcing. Adopting the same approach, Li et al. (2008) showed the largest depth-integrated body force is 4.0 m$^2$s$^{-2}$ and they suggested that the internal tide generated in the Luzon Strait can propagate 880 km without nonlinear disintegration in the deep basin of the SCS, reach the continental slope to the northeast of Hainan Island and then evolve into ISWs of depression. Similarly, because the largest value in the present computation is 1.44 m$^2$s$^{-2}$ (Fig. 9), we deduce that the internal tide generated in the Luzon Strait may also be able to propagate 520 km without nonlinear disintegration in the deep basin, reach the continental slope near the S7 station and then evolve into convex ISWs. More important is that the long-range propagation of internal tides without nonlinear disintegration in the deep basin has been validated by the field measurement, which shows a distinct lack of nonlinear internal waves during the winter in 2005 (Ramp et al.2010). As for the local continental slope shelf, Fig.8 shows negligible tidal forcing, consistent with the modeling study that shows that the baroclinic tidal energy conversion rate for the $M_2$ constituent during the winter is only 5 percent of that during the summer (Wu et al, 2013), demonstrating that the semidiurnal IT generated at the local continental slope shelf can be ignored. Moreover, another fact supporting the disregard of the contribution of the local IT is that $F > 0.25 \text{m}^2\text{s}^2$ is usually regarded as a critical value for the generation of ISWs (Lozovatsky et al., 2012; Bai et al., 2014). As a result, we can conclude that strong ITs could have been generated in the Luzon Strait on 23 December, 2005 and subsequently arrived at the upper continental slope two or three days later, thereby contributing to the outstanding records of convex ISWs at the mooring position reported in Yang et al.(2009) .

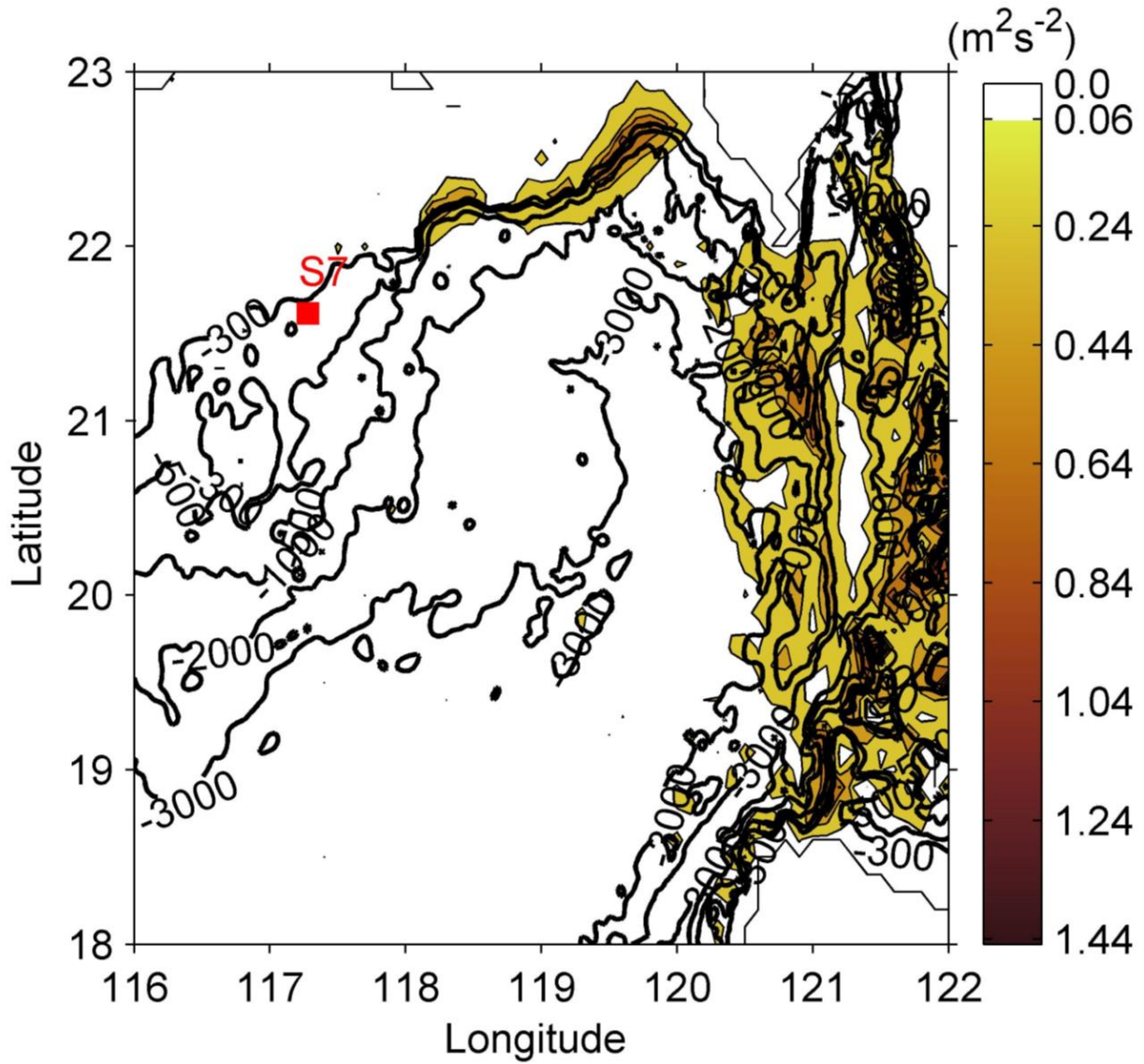

Fig. 8 Map of the maximum depth-integrated barotropic tidal forcing on 23 December 2005. The red square is mooring position S7 as reported in Yang et al. (2009)

**6.2 Implication of the Study**

We demonstrate a new possible generation mechanism for convex ISWs during the winter in the northern SCS, which are generated by the nonlinear transformation of a shoaling mode-2 semidiurnal IT, with the high possibility of originating from the Luzon Strait. The finding provides an alternative perspective on the generation of mode-2 internal waves and further, a theoretical reference for in situ experiments. It is well known that in situ experiments are crucial and expensive in the correctly understanding the ocean. However, having a certain understating of the questions that are to be solved is a necessary premise to conduct a relevant in situ experiment. The demonstrated generation mechanism could stimulate oceanographers to conduct more in situ experiments to examine its validity. For example, two moorings at depths of about 1000 m and 350 m on the continental slope may be deployed to locate the emergence and reveal more precise physics of second mode ISWs based on the present finding. Furthermore, taking advantage of these in situ measurements, accurate information about incoming ITs and varying background stratification can be input into the present numerical model or an ocean model such as the MITgcm, and thus detailed physics regarding to the mode-2 internal waves will be available.

In addition, it is well known that mode-1 and mode-2 ITs dominate the magnitude and geography of mixing in the open ocean (Alford and Zhao, 2007). According to the modeling results in the middle and lower panels in Fig. 6, strong energy transfer from mode-1 to mode-2 could also potentially occur in the open ocean. Thus, considering intermodal coupling shall lead to a more detailed understanding of the energy budget of low-mode ITs within the ocean interior.

## 7. Conclusions

The in situ measurements of Yang et al. (2009) showed a lot of interesting and valuable information of mode-2 internal waves during the winter on the continental slope of the northern SCS. In the light of their in situ measurements, we focus on explaining the generation of convex ISWs reported during the winter on the upper continental slope of the northern SCS. A multi-modal equation set is used to conduct numerical tests. Forced by typical types of ITs, the numerical results are in agreement with both theoretical predictions and field observations.

The study indicates that the field-observed convex ISWs were generated by a shoaling semidiurnal mode-2 IT that originated from the Luzon Strait. The IT generated two or three groups of convex ISWs within a semidiurnal tidal period that were dependent on the amplitudes of ITs. However, when the amplitudes of the ITs are small, no convex ISWs will emerge.

The strong scattering of a mode-1 IT into mode-2 waves also occurred, leading to the formation of a low-mode IT and a nonlinear mode-2 wave.

**Appendix A: In situ measurements of convex ISWs reported by Yang et al. (2009)**

The consecutive observations of convex ISWs from 24 to 27 December, 2005 are shown in Fig. A1. It is a direct copy of Fig. 5 in the paper of Yang et al. (2009). It can be seen that, convex ISWs dominate in the vertical temperature profiles and there is a deep thermocline which is located at 150-200 m. Yang et al. (2009) argued that the deep thermocline favors the generation of convex ISWs. More information about the characteristics and generation of convex ISWs shall be referred to Yang et al. (2009).

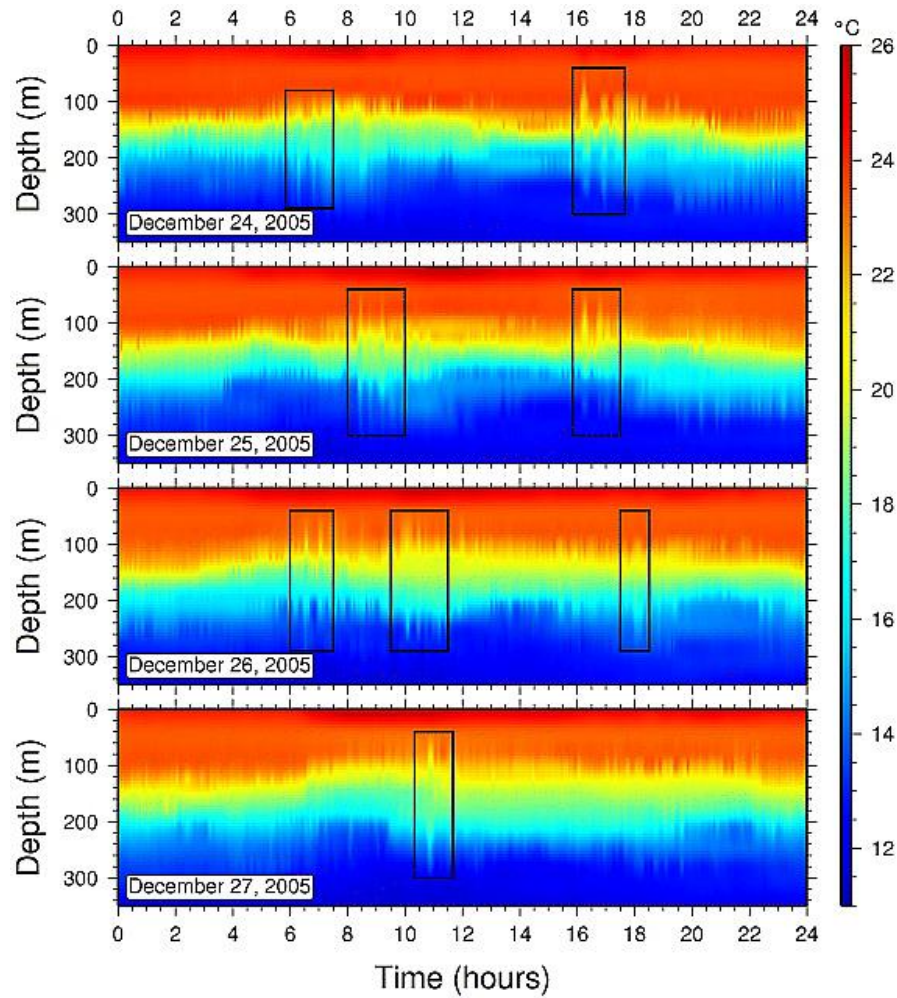

Fig. R1 Contour plots of the isothermal depth from 24 to 27 December 2005. The black rectangles indicate convex ISWs (adopted from Yang et al. 2009)

**Acknowledgments, Samples, and Data**